\newcommand{\Ni}{\mbox{$^{56}$Ni}}
\newcommand{\Co}{\mbox{$^{56}$Co}}

\newcommand{\kms}{\mbox{$\rm{km}\,s^{-1}$}}

\newcommand{\grizy}{\ensuremath{grizy_{\rm P1}}}

\newcommand{\msun}{\mbox{M$_{\odot}$}}

\newcommand{\Fe}{\mbox{$^{56}$Fe}}

\newcommand{\NaI}{Na\,{\sc i}}

\documentclass[twocolumn]{aastex631}

\newcommand{\thisSN}{\mbox{SN\,2023zaw}}
\newcommand{\SNxx}[1]{\mbox{SN\,#1}}

\newcommand{\tardis}{\textsc{tardis}}
\newcommand{\eg}{\mbox{e.g.}}
\newcommand{\ie}{\mbox{i.e.}}

\usepackage{booktabs,threeparttable}
\usepackage{amsmath}
\usepackage{rotating}
\usepackage{xcolor}

\shorttitle{SN\,2023zaw}
\shortauthors{Moore et al.}

\graphicspath{{./}{Figures/}}

\begin{document}

\title{SN\,2023zaw: the low-energy explosion of an ultra-stripped star}

\correspondingauthor{T.~Moore}
\email{tmoore11@qub.ac.uk}

\author[0000-0001-8385-3727]{T.~Moore}  
\affil{Astrophysics Research Centre, School of Mathematics and Physics, Queen's University Belfast, BT7 1NN, UK}
\affil{European Southern Observatory, Alonso de C\'{o}rdova 3107, Casilla 19, Santiago, Chile}
\author[0000-0002-8094-6108]{J.~H.~Gillanders}   
\affil{Astrophysics sub-Department, Department of Physics, University of Oxford, Keble Road, Oxford, OX1 3RH, UK}
\author[0000-0002-2555-3192]{M.~Nicholl}   
\affil{Astrophysics Research Centre, School of Mathematics and Physics, Queen's University Belfast, BT7 1NN, UK}
\author{M.~E.~Huber}  
\affiliation{Institute for Astronomy, University of Hawai'i, 2680 Woodlawn Drive, Honolulu, HI 96822, USA}
\author[0000-0002-8229-1731]{S.~J.~Smartt}  
\affil{Astrophysics sub-Department, Department of Physics, University of Oxford, Keble Road, Oxford, OX1 3RH, UK}
\affil{Astrophysics Research Centre, School of Mathematics and Physics, Queen's University Belfast, BT7 1NN, UK}
\author[0000-0003-4524-6883]{S.~Srivastav}  
\affil{Astrophysics sub-Department, Department of Physics, University of Oxford, Keble Road, Oxford, OX1 3RH, UK}
\author{H.~F.~Stevance}   
\affil{Astrophysics sub-Department, Department of Physics, University of Oxford, Keble Road, Oxford, OX1 3RH, UK}
\affil{Astrophysics Research Centre, School of Mathematics and Physics, Queen's University Belfast, BT7 1NN, UK}
\affil{Department of Physics, The University of Auckland, Private Bag 92019, Auckland, New Zealand}
\author[0000-0002-1066-6098]{T.-W.~Chen}
\affil{Graduate Institute of Astronomy, National Central University, 300 Jhongda Road, 32001 Jhongli, Taiwan}
\author{K.~C.~Chambers}  
\affil{Institute for Astronomy, University of Hawai'i, 2680 Woodlawn Drive, Honolulu, HI 96822, USA}

\author[0000-0003-0227-3451]{J.~P.~Anderson}   
\affil{European Southern Observatory, Alonso de C\'{o}rdova 3107, Casilla 19, Santiago, Chile}
\affil{Millennium Institute of Astrophysics (MAS), Nuncio Monse\~{n}or S\'{o}tero Sanz 100, Providencia, Santiago, Chile}
\author[0000-0003-1916-0664]{M.~D.~Fulton}   
\affil{Astrophysics Research Centre, School of Mathematics and Physics, Queen's University Belfast, BT7 1NN, UK}

\author[0000-0001-9309-7873]{S. R. Oates}
\affiliation{Department of Physics, Lancaster University, Lancaster, LA1 4YB, UK}

\author{C.~Angus}  
\affil{Astrophysics Research Centre, School of Mathematics and Physics, Queen's University Belfast, BT7 1NN, UK}
\author{G.~Pignata}  
\affil{Instituto de Alta Investigación, Universidad de Tarapacá, Arica, Casilla 7D, Chile}

\author[0000-0002-9986-3898]{N.~Erasmus}  
\affil{South African Astronomical Observatory, PO Box 9, Observatory 7935, Cape Town, South Africa} 
\affil{Department of Physics, Stellenbosch University, Stellenbosch, 7602, South Africa}

\author[0000-0003-1015-5367]{H.~Gao} 
\affil{Institute for Astronomy, University of Hawai'i, 2680 Woodlawn Drive, Honolulu, HI 96822, USA}
\author{J.~Herman} 
\affil{Institute for Astronomy, University of Hawai'i, 2680 Woodlawn Drive, Honolulu, HI 96822, USA}

\author{C.-C.~Lin}   
\affil{Institute for Astronomy, University of Hawai'i, 2680 Woodlawn Drive, Honolulu, HI 96822, USA}
\author{T.~Lowe}
\affiliation{Institute for Astronomy, University of Hawaii, 2680 Woodlawn Drive, Honolulu, HI 96822}
\author[0000-0002-7965-2815]{E.~A.~Magnier}  
\affiliation{Institute for Astronomy, University of Hawaii, 2680 Woodlawn Drive, Honolulu, HI 96822}

\author{P.~Minguez}   
\affiliation{Institute for Astronomy, University of Hawaii, 2680 Woodlawn Drive, Honolulu, HI 96822}
\author[0000-0001-8771-7554]{C.-C.~Ngeow}   
\affil{Graduate Institute of Astronomy, National Central University, 300 Jhongda Road, 32001 Jhongli, Taiwan}
\author{X.~Sheng}  
\affil{Astrophysics Research Centre, School of Mathematics and Physics, Queen's University Belfast, BT7 1NN, UK}

\author[0000-0002-9774-1192]{S.~A.~Sim}  
\affil{Astrophysics Research Centre, School of Mathematics and Physics, Queen's University Belfast, BT7 1NN, UK}

\author{K.~W.~Smith}  
\affil{Astrophysics Research Centre, School of Mathematics and Physics, Queen's University Belfast, BT7 1NN, UK}

\author{R.~Wainscoat}   
\affiliation{Institute for Astronomy, University of Hawaii, 2680 Woodlawn Drive, Honolulu, HI 96822}

\author{S.~Yang}   
\affil{Henan Academy of Sciences, Zhengzhou 450046, Henan, China}
\author[0000-0002-1229-2499]{D.~R.~Young} 
\affil{Astrophysics Research Centre, School of Mathematics and Physics, Queen's University Belfast, BT7 1NN, UK}
\author{K.-J.~Zeng}   
\affil{Graduate Institute of Astronomy, National Central University, 300 Jhongda Road, 32001 Jhongli, Taiwan}

\begin{abstract}
Most stripped-envelope supernova progenitors are thought to be formed through binary interaction, losing hydrogen and/or helium from their outer layers. Ultra-stripped supernovae are an emerging class of transient which are expected to be produced through envelope-stripping by a NS companion. However, relatively few examples are known and the outcomes of such systems can be diverse and are poorly understood at present. Here, we present spectroscopic observations and high-cadence, multi-band photometry of SN\,2023zaw, a rapidly evolving supernova with a low ejecta mass discovered in a nearby spiral galaxy at D = 39.7\,Mpc. It has significant Milky Way extinction, $E(B-V)_{\rm MW} = 0.21$, and significant (but uncertain) host extinction. Bayesian evidence comparison reveals that nickel is not the only power source and an additional energy source is required to explain our observations. Our models suggest an ejecta mass of  $M_{\rm ej} \sim 0.07$\,\msun\ and a synthesized nickel mass of $M_{\rm Ni} \sim 0.007$\,\msun\ are required to explain the observations. We find that additional heating from a central engine, or interaction with circumstellar material can power the early light curve.  

\end{abstract}


\section{Introduction} \label{sec:intro}
Modern sky surveys such as the Asteroid Terrestrial-impact Last Alert System (ATLAS; \citealt{2018PASP..130f4505T}), Zwicky Transient Facility (ZTF; \citealt{2019PASP..131a8002B}), and the Panoramic Survey Telescope and Rapid Response System  (Pan-STARRS; \citealt{Chambers2016}) are revealing the extremes of core-collapse supernovae (SNe) and optical transients \citep{2019NatAs...3..697I,2019NatAs...3..717M}. A small number of SNe, often belonging to the hydrogen poor Types Ib and Ic, show rapid evolution and brighten and fade on timescales much faster than typical classes of SNe \citep{2010Sci...327...58P, 2013ApJ...774...58D,2018Sci...362..201D,2018ApJ...865L...3P,2020A&A...635A.186P,2020ApJ...889L...6C,2023ApJ...959L..32Y, 2023ApJ...949..120H}. Generally, a small ejecta mass is invoked to explain the rapid evolution of fast transients \citep{2017MNRAS.466.2085M}. A small ejecta mass reduces the photon diffusion timescale, allowing the light curve to peak and begin to decline rapidly. 
Low ejecta mass interpretations require a physically compatible powering source. Invoking radioactive \Ni\ in fast-evolving SNe frequently produces unphysical ejecta mass to nickel mass ratios \citep{2018ApJ...865L...3P,2020A&A...635A.186P,2020MNRAS.497..246G,2020ApJ...889L...6C}.Additional mechanisms have been suggested to boost the luminosity of these supernovae; e.g.\ interaction with circumstellar material, or energy injection from a magnetar \citep{2020ApJ...900...46Y,2022ApJ...927..223S}.

In this paper, we present spectrophotometric follow-up observations of the rapidly evolving \thisSN.\footnote{While preparing this manuscript, another pre-print on the same source appeared on the arXiv \citep{2024arXiv240308165D}.} Classified as a Type Ib SN, \thisSN\, rises rapidly to maximum light ($<$ 4 days) followed by a similarly fast decline, comparable to the fast fading Type~I SN 2019bkc \citep{2020ApJ...889L...6C, 2020A&A...635A.186P}. We compare a range of physical models by fitting semi-analytical models, we show that \thisSN\, cannot be powered solely by \Ni\, decay, the normal power source of core-collapse SNe. 

\section{Discovery and Follow-up}\label{sec:disc_follow}

\thisSN\ was  discovered on 2023 December 7 05:34~UTC (MJD 60285.23) by ZTF 
\citep{2019PASP..131a8002B} and registered on the Transient Name Server at 11:50~UTC \citep{2023TNSTR3158....1S} with the discovery mag $g = 19.34$. All observational phases in this section are quoted in observer-frame days, relative to the ZTF discovery epoch. We independently detected \thisSN\,in ATLAS data \citep{2020PASP..132h5002S}
a few hours later at 08:00 UTC
as the field visibility moved from California to Hawaii, at mag $o = 18.74$. 
The transient is offset 8.97" N, 19.15" W from UGC 03048, a spiral galaxy with a redshift from the NASA Extragalactic Database (NED) of 0.010150 $\pm$ 0.000026 \citep{2005ApJS..160..149S}. From NED the median redshift-independent distance to UGC 03048 is 39.7\,Mpc, based on the Tully-Fisher method \citep{2013AJ....146...86T}. \thisSN\ is located on the edge of one of the two prominent arms of UGC 03048. The Milky Way extinction along this line of sight is \mbox{A$_{\rm V}$ = 0.66 mag} \citep{2011ApJ...737..103S}. \NaI\ D lines in the classification spectrum suggest additional host extinction is significant \citep{2012MNRAS.426.1465P}.

Four AstroNotes regarding \thisSN\ were released on the Transient Name Server\footnote{\url{https://www.wis-tns.org/object/2023zaw}} at the time of discovery, commenting on its early evolution. \citet{2023TNSAN.335....1K} highlighted the discovery and fast fading nature of \thisSN, along with an observation of the transient with NOT/ALFOSC. The Kinder project \citep{2023TNSAN.338....1L} reported a color-dependent fade using observations performed on the 40-cm SLT at Lulin Observatory, Taiwan. In \citet{2023TNSAN.339....1F}, we reported the combined ATLAS and ZTF data and highlighted that this source was flagged by our `Fastfinder' filter and annotator on the Lasair broker\footnote{\url{https://lasair-ztf.lsst.ac.uk}} \citep{2019RNAAS...3a..26S} to find fast-evolving objects in the ZTF public alert stream. Both  \citet{2023TNSAN.335....1K} and \cite{2023TNSAN.339....1F} identified \thisSN\, as a fast-fading, sub-luminous and red transient. Spectroscopic observations with Keck \citep{2023TNSAN.340....1K}  reported an apparent similarity with the candidate `.Ia' SN 2010X \citep{2010ApJ...723L..98K}. Finally, \citet{2023TNSAN.341....1G} classified the object as a Type Ib SN based on observations performed with Gemini-N/GMOS, and this spectrum was immediately made public on the TNS. 

\subsection{Photometry}

Photometry for \thisSN\ (internal name ATLAS23wuw) were obtained from the ATLAS forced photometry server \citep{2021TNSAN...7....1S} and binned by day. The ATLAS \citep{2018PASP..130f4505T} system is an all-sky survey for potentially dangerous near-Earth objects.  ATLAS data are processed using the ATLAS Science Server \citep{2020PASP..132h5002S} to search for stationary transients.
We obtained measurements in the $g$ and $r$-bands using the Lasair broker \citep{2019RNAAS...3a..26S} and public ZTF stream data.\footnote{\url{https://lasair-ztf.lsst.ac.uk/objects/ZTF23absbqun}}

We triggered follow-up observations with the 1.8-m Pan-STARRS1 telescope on the Haleakala mountain, Hawaii \citep{Chambers2016}. The Pan-STARRS1 telescope has a 7\,deg$^2$\ field of view and features a 1.4~gigapixel camera. High-cadence observations in the \grizy-bands were taken from +6 to +44 days post-discovery. 
 Optical imaging was triggered with the 2.0-m Liverpool Telescope \citep[LT;][]{2004SPIE.5489..679S} using IO:O in $riz$ bands under program PL23B26 (PI: M. Fulton). Measurements were made by PSF fitting using Source-Extractor \citep{1996A&AS..117..393B} with local background subtraction.
We observed \thisSN\ with the 0.4-m SLT telescope as a part of the Kinder project \citep{2021TNSAN..92....1C} and measured PSF $griz$-band photometry. 
Three epochs of photometric observations were performed with the GMOS-N instrument at the Gemini-North \mbox{8.1-m} telescope, under program ID \mbox{GN-2023B-Q-125} (PI: M.~Huber). Photometric Gemini observations were obtained at phases of +44\,d ($riz$-band), +54\,d ($riz$-band) and +56\,d ($ri$-band) post-discovery.
These observations were bias-subtracted and flat-field corrected using standard recipes in \texttt{DRAGONS} \citep{Labrie2023_DRAGONS, DRAGONS_zenodo}. We also present three epochs of $r$-band photometry derived from the acquisition images obtained prior to our spectroscopic observations with GMOS-N (see Section~\ref{sec:spectroscopy} for details). Aperture photometry was performed using \texttt{PSF} \citep[][]{2023ApJ...954L..28N} with a small optimized aperture, an encircled energy correction, and local background subtraction.
The Ultra-Violet and Optical Telescope \citep[UVOT;][]{2005SSRv..120...95R} onboard the \textit{Neil Gehrels Swift Observatory} \citep[\textit{Swift};][]{2004ApJ...611.1005G} satellite observed \thisSN. A single $uvm2$ exposure was taken at +6~days and followed up in the $u$, $b$, $v$, $uvw1$, $uvm2$ and $uvw2$ bands at two epochs +8 days to +11 days from discovery. The images at each epoch were co-added, and the count rates obtained from the stacked images using the {\it Swift} tool {\texttt{uvotsource}}. To extract the source counts, we used a source aperture of $5 \arcsec$ radius and an aperture of $20 \arcsec$ radius for the background. The source count rates were converted to magnitudes using the UVOT photometric zero points \citep{poole,bre11}. All \textit{Swift} observations are non-detections of the transient.

\begin{figure}
    \centering
    \includegraphics[width=\linewidth]{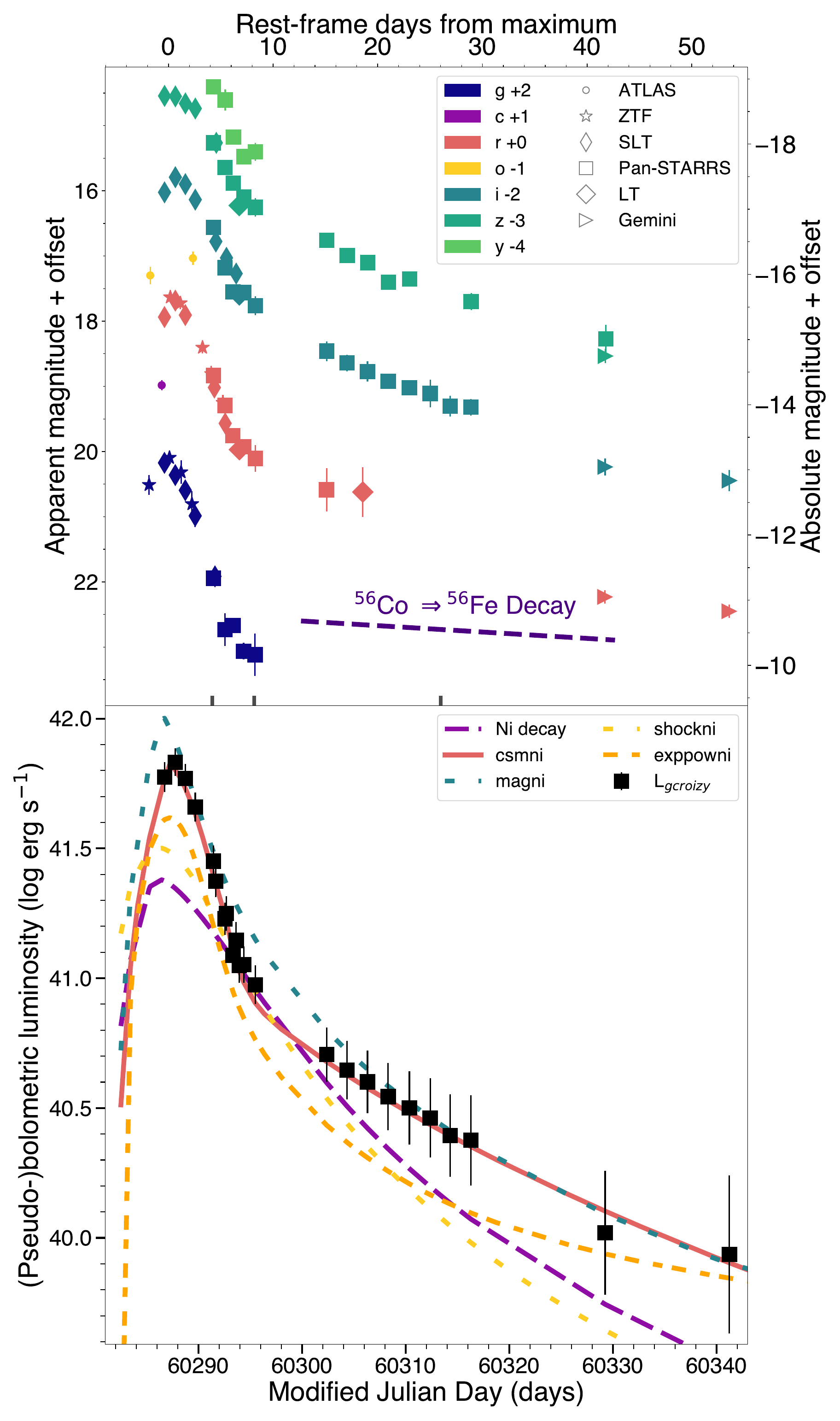}
    \caption{
        Top: Multicolor light curves with corrections for Milky Way foreground extinction and time dilation (for $z = 0.010150$) applied. Each telescope is shown with a different marker, and unfilled markers indicate upper limits. We exclude {\it Swift}/UVOT non-detections for visual clarity. Additionally, we show the expected decline rate of a \Co\ tail \citep[0.98 mag / 100 day;][]{1989ApJ...346..395W}. Bottom: A pseudo-bolometric (\textit{gcroizy}-band) light curve of \thisSN\, compared to the model light curves from Section \ref{sec:lightcurve_model} integrated between the \textit{gcroizy}-band observations. Model light curves are corrected for foreground extinction and the modeling derived estimate of the host extinction.}
    \label{fig:phot}
\end{figure}

The Milky Way extinction-corrected light curve of \thisSN\ is presented in Figure \ref{fig:phot} (see Section \ref{sec:host_extinction} for details of extinction estimation). We also present a pseudo-bolometric light curve calculated using the public \texttt{SUPERBOL} code \citep[see][for a description]{2018RNAAS...2..230N}. \texttt{SUPERBOL} applies a simple interpolation to the light curve and integrates underneath the observed photometric (\textit{gcroizy}-band) observations. The error bars on $L_{\rm bol}$ after $t=20$ days are significantly larger than the errors on the individual photometric data points. This accounts for the fact the SED is not well sampled in wavelength by the photometry (having only 2 or three bands) and the uncertainty in the effective temperature is propagated into the bolometric luminosity estimate (see \citet{2018RNAAS...2d.230N} for details).
All photometry presented in this work has been made available as a machine-readable table.

\subsection{Spectroscopy} \label{sec:spectroscopy}

We observed \thisSN\ at three different phases with the Gemini-North/GMOS-N instrument under program ID \mbox{GN-2023B-Q-125} (PI: M.~Huber).
Our three observations were taken at +6, +10 and 28 days post-discovery
(corresponding to phases from maximum light of $\approx +4.2$, +8.2 and +26.0~days, respectively).
All observations were performed using the R400 grating, sampling the $\approx 4200 - 9100$\,\AA\ wavelength range at a spectral resolution of R $\sim 1000$ for the 1\arcsec\ slit-width employed.
All three epochs of Gemini observations were reduced using the \texttt{DRAGONS} pipeline \citep{Labrie2023_DRAGONS, DRAGONS_zenodo} following standard recipes, and the spectra were all flux-calibrated against the same standard star. The contribution of the host galaxy was estimated and subtracted, and each reduced, co-added spectrum agrees well with the background-subtracted Pan-STARRS photometry obtained at the same epoch.
All spectra in this work will be made publicly available on the WISeREP repository \citep{2012PASP..124..668Y}.

\section{Analysis}
\label{sec:analysis}
\subsection{Host Galaxy and Milky Way Foreground Extinction} \label{sec:host_extinction}

There is a strong and narrow absorption line in the +4.2 and +8.2 day GMOS-N spectra, consistent with \NaI~D absorption at the redshift of UGC~03048. The GMOS-N spectral resolution does not allow the D$_1$ and D$_2$ components to be separately measured. After normalizing the spectrum, we fit a single Gaussian to the blended absorption line, finding a center $\lambda_c = 5953.94$\,\AA\ ($z = 0.0104$), FWHM $= 11.4$\,\AA, and an equivalent width, \mbox{${\rm EW} = 2.10 \pm 0.22$\,\AA}. At this redshift, the \NaI~D lines are separated by 6.03\,\AA, and the expected instrumental width of a single line is FWHM~$\simeq 6$\,\AA, giving an expected instrumental width for the unresolved \NaI~D blend of FWHM~$\simeq 9$\,\AA.

Measurements of the equivalent width of the \NaI\ doublet have been shown to be correlated with the line-of-sight extinction \citep{2012MNRAS.426.1465P}, and this method has often been applied to extragalactic transients. While there is a reasonably linear relation between line strength and $E(B-V)$, up to a total \mbox{${\rm EW}_{\rm (D_1 + D_2)} \simeq 0.7$\,\AA}, the relationship then saturates. No quantitative and unique measurement of $E(B-V)$ appears
possible beyond this, but we can say that an ${\rm EW} = 2.10 \pm 0.22$\,\AA\ requires a minimum of $E(B-V)_{\rm host} \gtrsim 0.5$ mag. The Milky Way foreground extinction is also significant along this line of sight, with $E(B-V)_{\rm MW} = 0.2$ mag \citep{2011ApJ...737..103S}. Throughout the rest of this manuscript, we apply a total extinction of $E(B-V)_{\rm total} = 0.7$ mag, noting that a somewhat higher value cannot be discounted. In Section\,\ref{sec:CSMNi} we show that  fitting the light curves with physical models and allowing the extinction prior to vary recovers a favored value compatible with our adopted value where R$_V$ = 3.1.

\subsection{Light Curves} \label{subsec:lc}

\begin{figure}
    \centering
    \includegraphics[width=\linewidth]{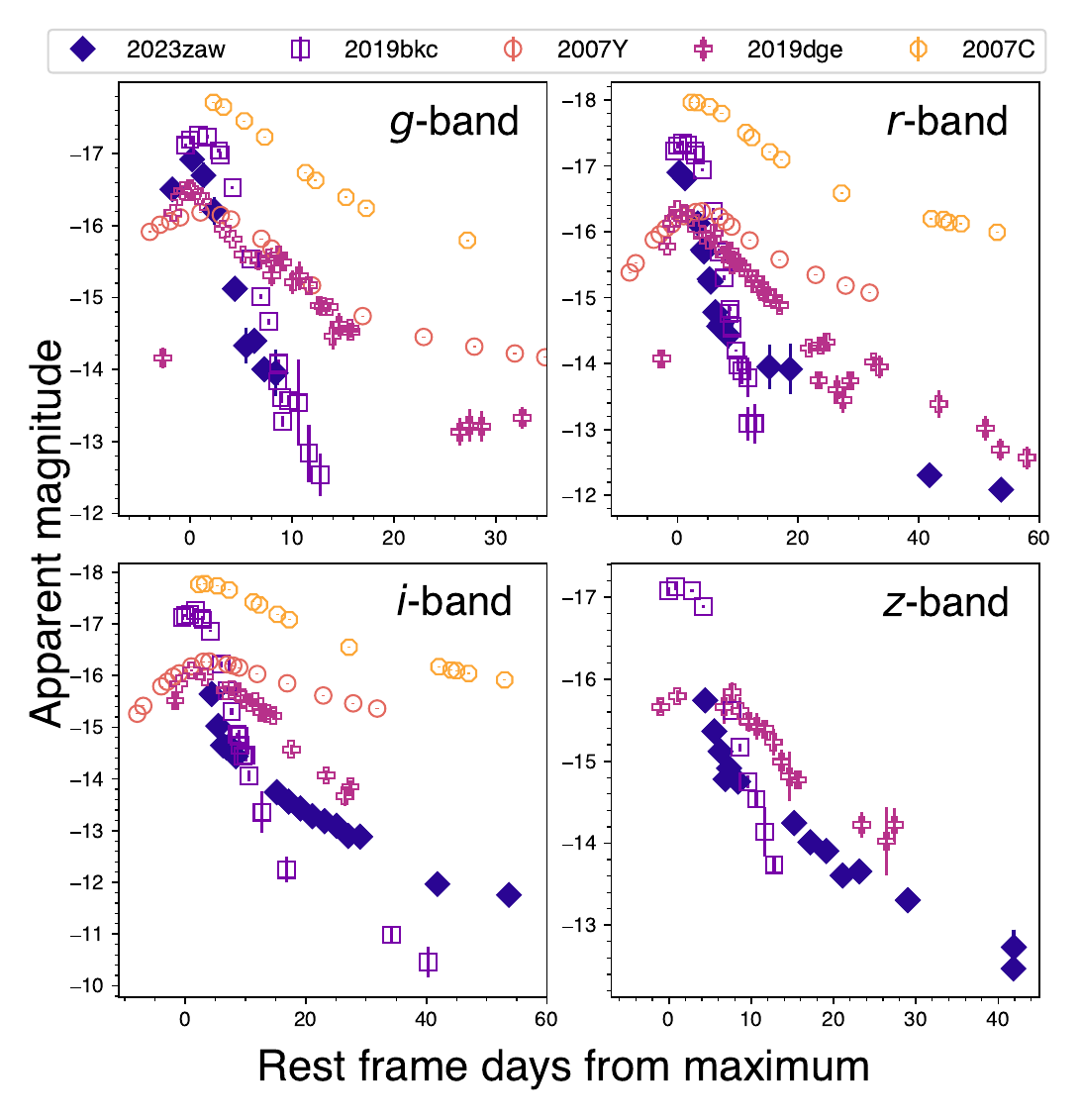}
    \caption{Light curves of \thisSN\, in the $griz$-bands compared to the fast supernovae SN 2019bkc \citep{2020A&A...635A.186P, 2020ApJ...889L...6C}, SN 2019dge \citep{2020ApJ...900...46Y}, and the representative Type SN Ib SNe 2007Y \citep{2009ApJ...696..713S} and SN 2007C \citep{2011ApJ...741...97D,2018A&A...609A.134S}. Each light curve is in the rest frame and has been corrected for Milky Way and host galaxy extinctions \citep{2011ApJ...741...97D}. These supernovae were chosen to represent the population of ultra-stripped SNe and typical Type Ib SNe. }
    \label{fig:compare}
\end{figure}

The light curves of \thisSN\ are shown in Figure~\ref{fig:phot}. Shortly after discovery, \thisSN\ reached a maximum brightness of r = 17.6 mag and g = 18 mag. Due to the short rise time we only observe the rising portion of the light curve in the $go$-bands.  We determine a time of maximum light of MJD $60287.1\pm0.2$ from fitting a polynomial to the ZTF $g$-band. The rapid rising phase of the light curve is not well observed, but the time from explosion to $g$-band peak is constrained to be less than four days by the ATLAS $o$-band non-detections (at depths corresponding to $M_{o}$ of $-14.5$ and $-14.8$ mag) at 2.7 and 1.8 days pre-discovery, respectively. The photometric evolution after maximum light is similarly rapid. \thisSN\, evolves extremely fast when compared to the representative Type Ib SN 2007Y \citep{2009ApJ...696..713S} in Figure \ref{fig:compare}. Initially \thisSN\ fades $\sim3$ magnitudes in $\sim10\,$days in the $r$-band, with SN 2007Y fading by less than a magnitude during the same interval. During this early rapid fade, \thisSN\ is comparable to SN 2019bkc, the fastest known Type I SN. By around 10 days after maximum, \thisSN\ settles to an apparent radioactive tail, which we observe in the $riz$-bands for approximately 40 rest-frame days. 

\begin{figure*}
    \centering
    \includegraphics[width=0.9\linewidth]{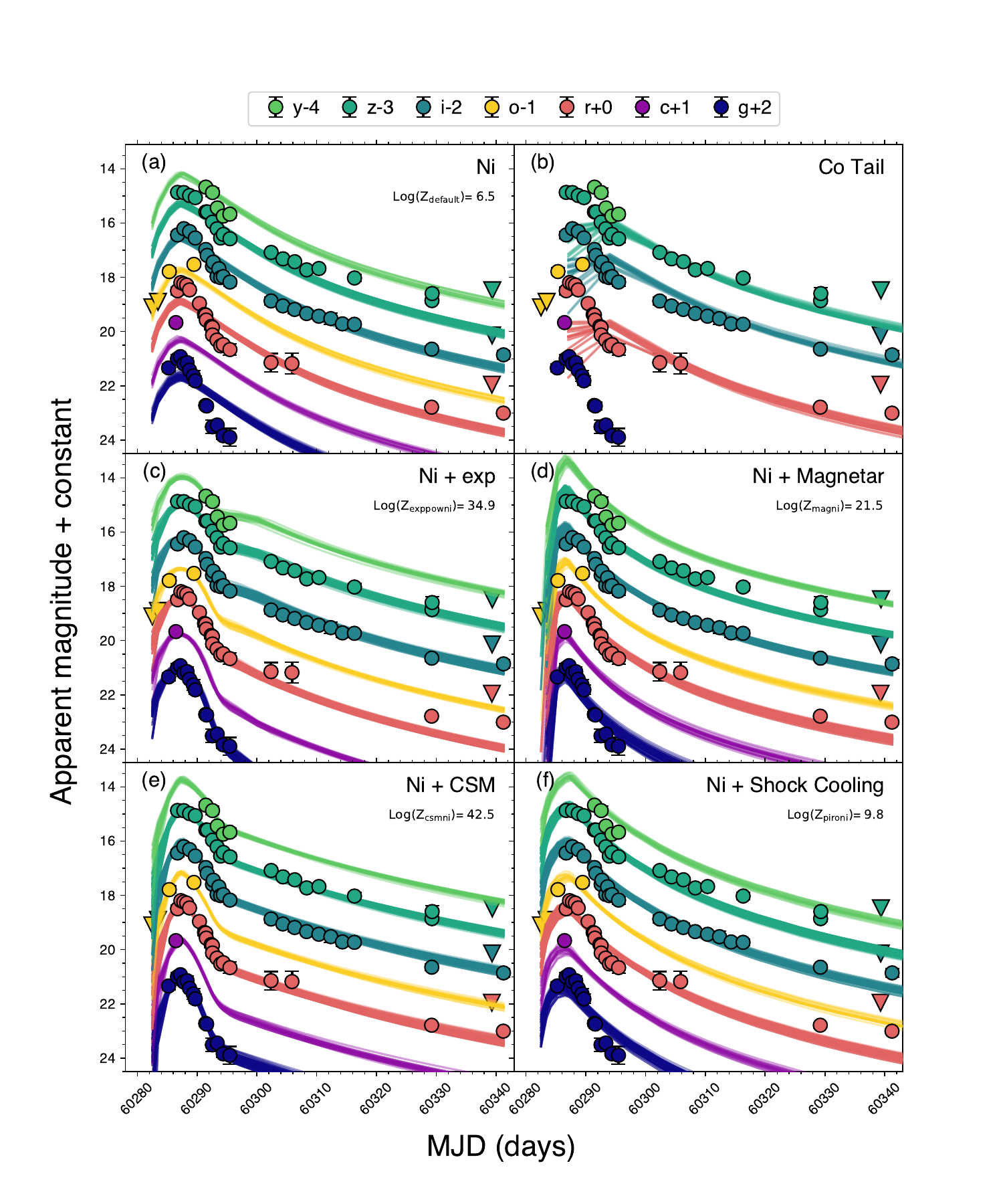}
    \caption{Results of the \texttt{MOSFiT} \citep{2018ApJS..236....6G} model analysis for each physical model compared to the complete light curves, where downward arrows are upper-limits, realizations are repented as solid lines, and band colors match those in Figure \ref{fig:phot}. Panel (a) shows nickel-only model realizations. Panels  (c), (d), and (e) show the nickel + exp model \citep{2018ApJS..236....6G}, magnetar + nickel model \citep{2017ApJ...850...55N}, and circumstellar material interaction + nickel \citep{2017ApJ...849...70V} models respectively. Panel (b) shows a fit to only the late-time light curve to estimate a synthesized nickel-mass, and Panel (f) shows our shock-cooling emission + nickel, the shock cooling model follows \citep{2021ApJ...909..209P}. The Bayesian evidence for each model is calculated for model comparison. When quoting the evidence for each model the name of each model in the \texttt{MOSFiT} code is used to ensure the reproducability of these results. Model evidence is not included in panel (b) as the model is only evaluated against the light curve tail. The \texttt{MOSFiT} input file will be made available as a data behind the figure file with the online article and all models are publicly available.}
    \label{fig:mosfit}
\end{figure*}

\subsection{Light Curve Modeling}
\label{sec:lightcurve_model}

We use the Modular Open Source Fitter for Transients \texttt{MOSFiT},\footnote{\url{https://github.com/guillochon/MOSFiT}} which is a \texttt{Python}-based modular code that evaluates a user-defined physical model directly against the multicolor observed light curves of transients.  A detailed description of the motivation, usage, and structure of the code can be found in \cite{2018ApJS..236....6G} and \citet{2017ApJ...850...55N, 2017ApJ...849...70V} provides thorough discussion of the modeling framework, simplifying assumptions and explains the magnetar, nickel and CSM interaction models. 

The code uses semi-analytical approximations for a range of luminosity sources (e.g., radioactive decay, magnetar spin-down, or CSM interaction). Radiative transfer through the ejecta uses the simplified approach of Arnett \citep{1982ApJ...253..785A}, enabling the model realizations to be evaluated rapidly for Bayesian fitting. The radius of the photosphere is calculated assuming a constant velocity at early times, followed by recession at a constant temperature once the ejecta cool to a critical value (see \citealt{2017ApJ...850...55N} for a full description). Finally an analytic model SED is calculated and converted to magnitudes using the filter transmission function for evaluation against the input data.

This last step is perhaps the biggest approximation, and worth discussing in some detail. When fitting analytic models to transient data, an assumption has to be made about the spectral energy distribution (SED) somewhere in the process. This could be on the data side (estimating the bolometric luminosity from filtered data before fitting) or on the model side (translating a model luminosity into bands of interest using a model SED). In the limit of exquisite multi-band data, the former approach is more reliable, as the bolometric luminosity can be well constrained from the data, and bolometric models are less subject to the uncertain radiative transfer that controls the emergent SED. However, for many real transients, observations cover only a limited portion of the SED, and not all bands are obtained at all times (for example, transients are often observed rising in only one or two bands before follow-up is triggered; at late times we often lose sensitivity in the blue bands first as transients cool). In this case, significant SED assumptions must be made on the data side if we want to construct a bolometric light curve (e.g. extrapolating in wavelength with an assumed blackbody SED, or extrapolating in time by assuming a constant temperature at early or late times).

The impact of these choices on subsequent model fits can be very difficult to quantify. For this reason, \texttt{MOSFiT} typically fits directly to multi-band data, assuming that the model emits an SED that can be described analytically -- usually a blackbody with time-variable temperature, calculated from the luminosity using the Stefan-Boltzmann law. Although this represents a major simplification of the true SED, in practice the same SED assumption is used whether we are constructing a bolometric light curve from limited bands or fitting to data in multiple bands. The advantage of the latter approach is that it removes the need to extrapolate in the time dimension, and in principle the model posteriors will reflect the uncertainties in the photospheric temperature. This makes it possible to model a wide range of transients, not only those with excellent multi-band data. Fitting in multi-band space provides additional color information that can help to constrain extinction, which is important for transients like \thisSN\ that have uncertain host extinction. A bolometric light curve analysis can also help in this regard, and to ensure consistency, \texttt{MOSFiT} also enables the user to output a bolometric light curve for any given fit, which can be compared to a bolometric light curve estimated from the data. In this study, we will fit to the multi-band data, but ensure we check for consistency in this way (see Figure \ref{fig:phot}).

In this work we use the \texttt{dynesty} \citep{2020MNRAS.493.3132S} nested sampling package in \texttt{MOSFiT} to evaluate posteriors for a series of different models. For all models we assume an optical opacity $\kappa = 0.1$ cm$^2$ g$^{-1}$ and a gamma-ray opacity $\kappa_\gamma = 0.027$ cm$^2$ g$^{-1}$. We apply a constraint to the host galaxy H column density of n$_{\rm H, host} > 3.4\times10^{21}\,\rm cm^{-2}$ for consistency with our adopted minimum host extinction $E(B-V)_{\rm host} \gtrsim 0.5$ (Section \ref{sec:host_extinction}), assuming R$_V$ = 3.1, using \citep{1989ApJ...345..245C} and the relation n$_{\rm H} (\rm cm^{-2}) / \rm A_V (\rm mag) = 2.21\times10^{21}$ \citep{2009MNRAS.400.2050G}. All models that are considered in this work are shown in Figure \ref{fig:mosfit} together with the observed light curves of \thisSN. We also show the bolometric light curves of these models compared to the estimated bolometric luminosity of \thisSN\ in Figure \ref{fig:phot}. From our models we estimate the values of important parameters (e.g. ejecta mass $M_{\rm ej}$ and nickel fraction $f_{\rm Ni}$); these parameters are quoted in the text. We re-emphasize that our model fits use simplified analytic prescriptions, and as with any models they include significant systematic errors that are difficult to quantify. Despite this caveat, these analytic models are thought to be reliable at the order-of-magnitude level, and our results are readily comparable to previous applications of \texttt{MOSFiT} in the SESN literature \citep{2017ApJ...850...55N, 2022ApJ...941..107G}.

The rising phase of \thisSN\ is poorly sampled compared to the peak and tail of the light curve, which makes it difficult to achieve a good fit to the early light curve. Following \citet{1982ApJ...253..785A} and \citet{2015MNRAS.450.1295W}, we use the observed rise time $t_{\rm r}$ and an estimate of the photospheric velocity $v_{\rm ph}$ to estimate an ejecta mass $M_{\rm ej}$. This approach assumes a homologously expanding and spherical ejecta, but provides a simple estimate of $M_{\rm ej}\sim 1/2 \cdot \beta \cdot  c/ \kappa \cdot v_{\rm ph} t_{\rm r}^2$ \citep{2015MNRAS.450.1295W} using the photon diffusion timescale. We assume $\beta =13.7$ \citep{1982ApJ...253..785A} though in reality this depends on the unknown ejecta density profile, $c$ is the speed of light, and $\kappa$ is the opacity. To approximate the rise time we adopt the midpoint between the last non-detection and the first detection as the explosion epoch (MJD 60284.3, which is $\sim 0.9$ days before detection) and the $g$-band peak for the time of maximum, this gives $t_{\rm r}\sim 2.75 \,$days. Using the photospheric velocity $v_{\rm ph}$ derived from spectral modeling in Section \ref{sec:spec}, this analysis estimates an ejected mass $M_{\rm ej}\sim 0.07$\,\msun. By allowing a $\pm$1 day uncertainty on the rise time our mass estimate could vary by up to $0.06$\,\msun. To capture this constraint we adopt a prior on the ejecta mass  $M_{\rm ej} = 0.07\pm0.06$ \msun\,for all models.

\subsubsection{Nickel Decay}
We first attempt to fit the multiband light curves with heating only from \Ni\ decay, using the built-in \texttt{default} \texttt{MOSFiT} model \citep{1994ApJS...92..527N,2018ApJS..236....6G}. We assume that the ejecta is spherical, with constant gamma-ray and optical capacities, and that the Ni is concentrated in the center of the SN, closely following \citet{1982ApJ...253..785A}. The details of the \texttt{MOSFiT} diffusion implementation are described in \citet{2017ApJ...850...55N}. Here we highlight that $1-e^{-\psi_{leak} t^{-2}}$ describes the fraction of the input energy which is thermalized, where the leakage parameter is  $\psi_{leak} = \frac{3 \kappa_{\gamma}M_{ej}}{4 \pi v_{ej}^2}$ \citep{2015ApJ...799..107W}, and $\kappa_\gamma = 0.027$ cm$^2$ g$^{-1}$. With this prescription, only a decreasing fraction of the \Ni\ decay energy is able to heat the ejecta as time increases. 

Despite a reasonable agreement with $i$-band observations, overall this model does not provide a satisfactory fit to the light curves of \thisSN\,. During the SN rise  the model conflicts with the deep ATLAS $o$-band limits, and fails to reproduce the maximum luminosity in the $gcr$-bands. The model clearly diverges from the late-time $rz$-band tail and only adequately matches the decline between MJD 60300 and MJD 60320. We find the most probable model ejects $M_{\rm ej} \simeq 0.06$\,\msun\ of material with a large \Ni\ fraction $f_{\rm Ni} \simeq 0.90$, and a reasonable SN-like ejecta velocity $v_{\rm ej} \simeq 6400$\,\kms. This implies $M_{\rm Ni} \simeq 0.05$\,\msun was synthesized in the explosion, this is lower than typical for Type Ib SNe \citep{2019A&A...628A...7A,2023ApJ...955...71R}.  More problematic is the very large nickel fraction, contradicting our observed spectra. An ejecta of mostly \Ni\ and its decay products should be dominated by iron-group absorption in the blue; this is not observed in the spectra of \thisSN. Additionally, a nickel yield of $M_{\rm Ni}\simeq0.05$\,\msun is difficult to rationalize in the context of the ultra-stripped SNe (e.g. see discussion from \citealt{2022ApJ...927..223S} on iPTF14gqr). The light curve tail begins approximately ten days after maximum. The decline-rate slows and follows a power-law decline, an evolution which is similar to the $i$-band tail of the fast-fading SN 2019bkc  \citep{2020ApJ...889L...6C,2020A&A...635A.186P} where this was attributed to radioactivity. 
Clearly, a \Ni-only model does not reproduce our observations, requires an unrealistic nickel-fraction, and can only adequately explain the light curve tail. Therefore, we exclude the scenario where nickel-decay is the only mechanism powering the light curve peak of \thisSN\, and seek an explanation with another mechanism in addition to nickel decay. 

Assuming the heating at \emph{late times} is powered only by decay of \Co\ to \Fe, we can estimate a synthesized nickel mass for \thisSN. We apply the same model as before but with a restriction to fit only the late time $riz$-band data (MJD $>$ 60300). The model likelihood is only evaluated against the late-time photometry, but we provide a prior constraining the explosion epoch between the last ATLAS non-detection (MJD 60282.51) and the first detection (MJD = 60285.23). The tail-only fit requires that $M_{\rm Ni} \simeq 0.006$\,\msun\ was synthesized and ejected to power this phase of the light curve. Seeking to verify our method, we also reanalyze the light curve tail in SN 2019bkc and find a nickel mass $M_{\rm Ni} \simeq 0.005$\,\msun, which is compatible with the $M_{\rm Ni} = 0.001 - 0.01$\,\msun\ estimated by \citet{2020ApJ...889L...6C}. We also perform a consistency check by fitting the late time bolometric tail using the method of \citet{2013arXiv1301.6766K,2019MNRAS.484.3941W}, finding a consistent nickel mass of $M_{\rm Ni} = 0.002 - 0.006$\,\msun.

Motivated by the poor agreement of the nickel decay model to the initial light curve but apparent agreement to the light curve tail, we consider a generalized early-time heating source plus a nickel decay tail. This approach aims to investigate the compatibility of the observed light curves with a nickel power source and an additional mechanism without placing assumptions on the exact nature of the additional source. Using \texttt{MOSFiT} we adopt a physics-agnostic analytical prescription for an exponentially rising additional energy source that declines from its maximum ($t_{peak}$) following a power-law (named \texttt{exppow}). It is described by
$L = L_{\rm scale} \cdot (1 - e^{-t / t_{peak}})^{\alpha} \cdot (t / t_{peak})^{-\beta} $, where $L_{\rm scale}$, $\alpha$, $\beta$, and $t_{peak}$ are free parameters. Combining the \texttt{default} and \texttt{exppow} models we created a new \texttt{MOSFiT} model called \texttt{exppowni}.

We fit the \texttt{exppowni} model as before using the same constraints on opacity and host galaxy extinction. The model realizations are shown in Figure \ref{fig:mosfit}. We see good agreement with observations and consistency with the ATLAS non-detections, the observed color, and late-time luminosities. We estimate a nickel mass of $M_{\rm Ni} \simeq 0.006$\,\msun, which is in agreement with the fit for only the light curve tail. It is clear that an additional luminosity source is required to simultaneously match the fast rise, the peak luminosity and the observed \Co\ tail. The parameterized nature of the \texttt{exppowni} model provides insight into the timescale and energetics of the additional luminosity source. The non-radioactive heating reaches $t_{peak}$ between $1.6-2.6$ days after explosion and dominates the luminosity during this phase with a scale luminosity $L_{\rm scale}$ $\sim 10^{42}$ erg s$^{-1}$.

\subsubsection{Circumstellar Material Interaction + Nickel}
\label{sec:CSMNi}
We next investigate interaction with nearby circumstellar material (CSM) as the possible extra energy source for \thisSN. CSM  interaction can produce unusual and rapidly evolving transients \citep[e.g.][]{2023ApJ...956L..31M,2023arXiv230316925K,2023A&A...673A..27N, 2022AAS...24023208P}. We use the \texttt{CSMNI} model in \texttt{MOSFiT} \citep{2017ApJ...849...70V,2020RNAAS...4...16J,2023ApJ...956L..31M} which combines the luminosity of \Ni\ decay and heating from shock propagation following an ejecta-CSM collision. The CSM interaction physics is implemented following the treatment of \citet{2013ApJ...773...76C}. We use the adapted model setup used by \citet{2023ApJ...956L..31M} and \citet{2023ApJ...956L..34S} where the onset of interaction is delayed until the ejected material reaches an inner CSM radius. 

We evaluate the \texttt{csmni} model against our observations and show the CSM model realizations in Figure \ref{fig:mosfit}. This model fits the light curve peak very well and matches the overall light curve evolution in all bands, including close agreement to the late-time tail. When compared to the pseudobolometric light curve we achieve excellent agreement to the observed data. The derived model parameters are:
 $M_{\rm ej} \simeq 0.069\,\msun$,
$f_{\rm Ni} \simeq 0.13$,
$r_{\rm csm} \simeq 63$\,AU,
$M_{\rm csm} \simeq 0.23$\,\msun,
where $M_{\rm csm}$ is the mass of the CSM material and $r_{\rm csm}$ is the CSM radius. The CSM mass and radius show strong degeneracy. The derived kinetic energy for this model is low at $E_k \sim10^{49}$\,erg. This model implies $M_{\rm Ni} \simeq 0.008$\,\msun, which is compatible with our estimate from fitting only the light curve tail. 

The CSM interaction treatment assumes that the progenitor star is embedded in a spherically symmetric CSM shell with a power-law density profile described by a single power-law index $s$,  where $s = 2$ is a wind-like CSM and $s = 0$ is a shell of constant density, see \citet{2017ApJ...849...70V} for a full model description. Treating it as a free parameter in our fit, we find $s \simeq 0.68$ which does not indicate a strong preference for a constant density shell or wind mass loss history.

 We note the +61 day spectrum of \thisSN\ \citep{2024arXiv240308165D} shows narrow helium lines, which are also consistent with circumstellar interaction. These results suggest significant late-stage mass loss consistent with the binary mass transfer scenario simulated for a Type Ib/c SN progenitor in \citet{2022ApJ...940L..27W}. 
We note that if we allow the prior on the extinction of the host galaxy to broadly vary, the model converges to a host 
$E(B-V) \simeq 0.6$ (R$_V$ = 3.1), which is similar to our adopted extinction estimate from Section\,\ref{sec:host_extinction}.

\subsubsection{Central Engine + Nickel}

As an alternative, we next investigate the feasibility of central engine heating as the additional energy source using the magnetar central engine + nickel (\texttt{magni}) model described by \citet{2017ApJ...850...55N} and \cite{2022ApJ...941..107G}. This combines the luminosity of magnetar spin-down \citep{2010ApJ...717..245K, 2010ApJ...719L.204W} with radioactive heating. This model reproduces the data significantly better than radioactive heating alone, as shown in Figure \ref{fig:mosfit}, and matches the peak but diverges from the $riz$-band tail. This model achieves agreement within the uncertainties to the pseudobolometric light curve in Figure \ref{fig:phot}.  Our derived \texttt{magni} model parameters are:
ejecta mass  $M_{\rm ej} \simeq 0.055$\,\msun,
nickel fraction $f_{\rm Ni} \simeq 0.08$, and $M_{\rm Ni} \simeq 0.004$\,\msun.
We find  central engine parameters of
$P_{\rm spin} \simeq 6.3$\,ms and
$B_{\perp} \simeq 0.19 \times 10^{14}$\,G. The B-field required from our models is consistent with the population of superluminous supernovae. However, the spin period, $P_{\rm spin}$, is towards the longer end of the range of $1 - 6$\,ms \citep[e.g.][]{2016ApJ...818...94K, 2017ApJ...850...55N}. This model Given the low ejecta mass, the region of the ejected material which is being heated by the central engine should be visible. Therefore, unless the oxygen is doubly ionized we would expect the W-shaped O II absorption lines to be present in the spectrum which are the characteristic signature of central engines in Type I superluminous supernovae \citep{2016MNRAS.458.3455M,2018ApJ...855....2Q}; unfortunately this region of the \thisSN\ spectrum was not observed. The extinction of the host galaxy to broadly vary, the model converges to a host 
$E(B-V) \simeq 0.6$, this is similar to the value returned by the \texttt{csmni} model and our adopted extinction estimate. 

\subsubsection{Shock Cooling + Nickel}

Finally, we consider the combination of \Ni\ and shock cooling emission following the analytical model from \citet{2021ApJ...909..209P}. In this scenario the progenitor star possesses an extended envelope into which energy is deposited by the SN blast wave; the envelope then radiates this energy as it cools. We have added this model to \texttt{MOSFiT} as the \texttt{pironi} model.
We show our model realizations in Figure\,\ref{fig:mosfit}. Our model requires $f_{\rm Ni} \simeq0.01$,
$M_{\rm ej} \simeq 0.04$\,\msun,
$M_{\rm env} \simeq 0.3$\,\msun, and $R_{\rm star} \simeq 10^{10.5}\, \rm cm$. However, the shock cooling model shows relatively poor agreement with the overall light curve of \thisSN. It contradicts the early ATLAS non-detections and does not reproduce the observed tail, where the model is underluminous. 

We note that this model has better agreement to observations if we instead adopt the host galaxy extinction estimate from \citet{2024arXiv240308165D}, $\rm A_{v, host} = 1.12$ mag, $E(B-V)$ = 0.36 (R$_V$ = 3.1). Evaluating the \texttt{pironi} model with this value of host extinction, we find overall similar parameter values: $f_{\rm Ni} \sim 0.009$, an envelope mass $M_{\rm env} \sim 0.3$\,\msun, an ejected mass $M_{\rm ej} \sim 0.03$\,\msun\ and a stellar radius $R_{\rm star} \sim 10^{10.3}\, \rm cm$.

\subsubsection{Model Comparison}

To compare the models we use the Bayesian model evidence (marginal likelihood) scores for each model. These are returned by \texttt{MOSFiT} when using the nested sampler \texttt{Dynesty} \citep{2020MNRAS.493.3132S}. All \texttt{MOSFiT} analysis in this work uses a nested sampling approach to enable comparison between models. From the Bayesian evidence scores, a Bayes factor (BF) can be used to compare two models. The BF comparing model $x$ and model $y$ is given by B $\equiv$ Z$_{\rm x}$/Z$_{\rm y}$, where Z$_{\rm i}$ is the Bayesian evidence for model i. A Bayes factor B $>$ 10 indicates a strong preference, and B $>$ 100 is considered definitive. 

We find a BF Z$_{\rm expowni}$/Z$_{\rm nickel}\ \sim 10^{21}$, where Z$_{\rm exppowni}$, Z$_{\rm nickel}$ are the evidence for their respective models, this shows evidence favoring an additional powering source. This statistical test and the poor agreement of the pure \Ni model with the light curve provides evidence that \thisSN\ requires an additional power source.  Comparing our Ni model and shock cooling model, we find Z$_{\rm pironi}$/Z$_{\rm nickel}\ \sim 10^{1}$, the shock cooling + Ni model is preferred over Ni alone.
Comparing between models in this way our analysis favors CSM + Ni model above all others. When comparing the alternative models we  calculate Z$_{\rm csmni}$/Z$_{\rm magni}$ $\sim10^{10}$, favoring CSM interaction over the spin-down of a newly born neutron star. However, we note that the \texttt{MOSFiT} \texttt{CSMNI} model is the most complex model evaluated in this work, and the most flexible, while likely oversimplifying the physics involved. While the CSM + Ni model reproduces the light curve tail luminosity better than any other model, we also note that the blackbody spectrum assumed in \texttt{MOSFiT} may be unreliable at this phase as the ejected material becomes transparent. 

Through \texttt{MOSFiT} modeling we have shown that the total emission of \thisSN\ cannot be explained with radioactivity alone, as the required $f_{\rm Ni} \simeq0.9$ is incompatible with the spectroscopic observations (discussed in Section~\ref{sec:spec}). We have found evidence that suggests another energy source is required which peaks early in the evolution of \thisSN\ and dominates the emission at this phase. We disfavor shock cooling emission as this additional source due to poor agreement to the multicolor photometry, in particular our ATLAS non-detections and to the pseudo-bolometric light curve. Our analysis favors CSM interaction + nickel, with a CSM mass that is large but not beyond theoretical predictions \citep{2022ApJ...940L..27W}, but cannot rule out a central engine + nickel model for \thisSN. We consider both models as viable explanations for \thisSN.

\subsection{Spectral Modeling and Analysis}\label{sec:spec}

\begin{figure}
    \centering \includegraphics[width=\linewidth]{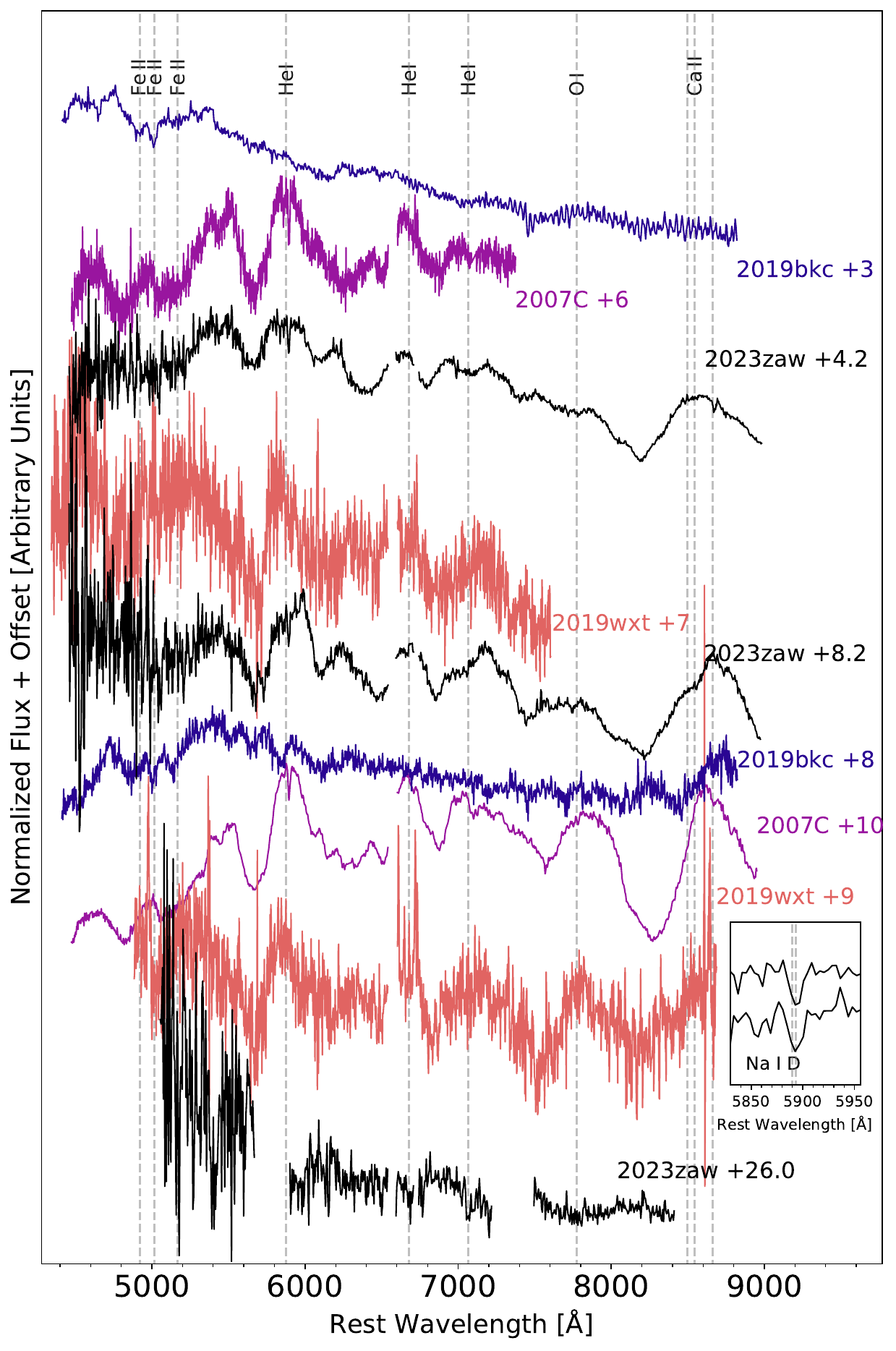}
    \caption{
        Optical spectroscopy of \thisSN . The phase in days relative to maximum light are indicated. These spectra have been telluric and Galactic line of sight extinction corrected, and corrected for host galaxy recessional velocity. The inset plot shows the blended Na I D lines in the +4.2 and +8.2~day spectra.
        We also include spectra from SN 2019wxt \citep{2023A&A...675A.201A}, SN 2007C \citep{2014AJ....147...99M}, and SN 2019bkc \citep{2020ApJ...889L...6C, 2020A&A...635A.186P} and include the approximate phase of each observation.
    }
    \label{fig:spec}
\end{figure}

We present post-peak spectra taken +4.2, +8.2 and +26.0~days after maximum light
in Figure \ref{fig:spec}. Our first spectrum (taken at +4.2\,d) shows well-developed absorption features and a prominent Ca II NIR triplet. Weak Fe II features may exist around 4500--5000\,\AA, and we find that a 5700\,K blackbody produces a continuum in agreement with the observed spectrum. As noted in Section \ref{sec:host_extinction}, the spectra show a prominent Na I D line blend. 

The spectra show little evolution between the +4.2 and +8.2 day observations, considering the rapidly evolving light curve. At +8.2\,d, the spectroscopic features have more developed line profiles and a broad emission feature at 7100\,\AA\ is apparent. Our final observation at +26.0\,d is mostly featureless and likely dominated by the host galaxy, therefore we exclude it from further quantitative analysis.

\begin{figure*}
    \centering
    \includegraphics[width=0.9\linewidth]{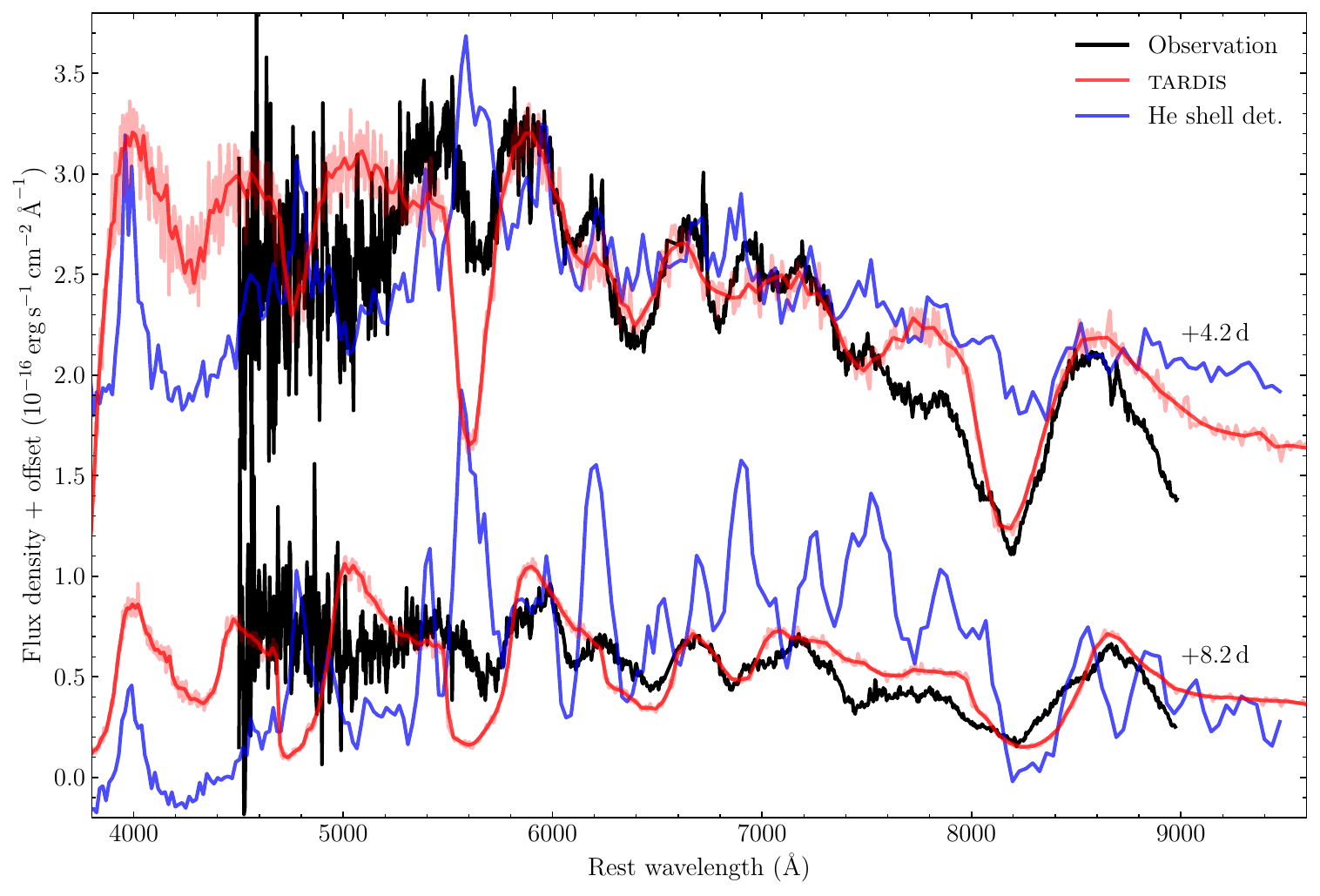}
    \caption{
        Best-fitting \tardis\ models (red), compared to the +4.2 and +8.2\,d observed spectra of \thisSN\ (black). The +4.2\,d observed spectrum (and associated model spectra) have been vertically offset for clarity (by \mbox{$6 \times 10^{-17} {\rm erg} \, {\rm s}^{-1} \, {\rm cm}^{-2} \, {\rm \AA}^{-1}$}). Also plotted are two model spectra from a helium shell detonation simulation (blue) presented by \cite{2012MNRAS.420.3003S}. These models have phases comparable to those of the observed spectra ($\Delta t < 0.13$\,d), and have been re-scaled and arbitrarily offset to roughly match the continua of the observed spectra.
    }
    \label{fig:TARDIS fits}
\end{figure*}

\thisSN\ has been proposed as a `.Ia' SN and compared to the .Ia SN candidate \SNxx{2010X} \citep{2023TNSAN.340....1K}.
.Ia SNe are the theorized explosion of a helium shell on the surface of a white dwarf \citep{2009ApJ...699.1365S, 2010ApJ...715..767S}. In Figure~\ref{fig:TARDIS fits} we present a spectroscopic comparison to He shell detonation models \citep{2012MNRAS.420.3003S}. Although the model provides some agreement to the SED of the observed spectrum at +4.2 days, it predicts emission at $\sim 5700$\,\AA, where in the observed spectrum we instead see strong He~I absorption. At +8.2\,d, the differences between the model and our observation become more stark, with the He detonation model showing strong emission features not present in the data. With such clear divergence from the He detonation model, we rule out a .Ia origin for \thisSN, in agreement with \citet{2024arXiv240308165D}.

As shown in Figure \ref{fig:spec}, \thisSN\ exhibits a striking similarity to \SNxx{2019wxt}, both in terms of the observed lines and overall SED shape. Here we undertake a similar process to that presented by \cite{Agudo2023_SN2019wxt}, using \tardis\ to model the photospheric-phase spectra.
\tardis\ \citep{Kerzendorf2014} is a one-dimensional, time-independent, Monte Carlo radiative transfer spectral synthesis code capable of simulating the spectra of an array of different explosive transients. 
Here we briefly describe the code, but for full details of our implementation, see \cite{Kerzendorf2014} and \cite{Agudo2023_SN2019wxt}. In brief, \tardis\ simulates the propagation of \textit{radiation packets} (analogous to photons) through some model ejecta structure, as defined by user inputs.
These packets undergo free $e^-$ scattering and bound--bound interactions with the ejecta material, and the ones
that emerge from the simulation are used to compute a synthetic spectrum, which is compared to observation \citep[usually through a visual `$\chi$-by-eye' approach; see \eg,][]{Stehle2005}. We then iteratively vary our input parameters to improve agreement between model and observation, until satisfactory agreement to the data has been reached.

While \tardis\ is a time-independent code, 
one can evolve the input parameters to obtain a sequence of self-consistent models, as we do here for the +4.2 and +8.2\,d spectra of \thisSN. These user-defined input parameters include specifying
the time since explosion, $t_{\rm exp}$ (which we set to be 2~days pre-maximum),
the inner and outer boundary of the computational domain (defined in velocity-space,
where $v_{\rm inner}^{+4.2\,{\rm d}} = 12500$\,km\,s$^{-1}$, $v_{\rm inner}^{+8.2\,{\rm d}} = 5000$\,km\,s$^{-1}$, and $v_{\rm outer} = 20000$\,km\,s$^{-1}$), the abundance and density of the ejecta material (here we use a uniform abundance across the entire ejecta and across both epochs; see Table~\ref{tab:TARDIS abundance}), and we invoke an exponential profile, where:

\begin{equation}
    \begin{split}
    \rho \left( v, t_{\rm exp} \right) = 2 \times 10^{-12} & \times \exp \left[\frac{- v}{6000 \ {\rm km\,s}^{-1}} \right] \\
    & \times \left( \frac{2 \ {\rm day}}{t_{\rm exp}} \right)^3 {\rm g\,cm}^{-3} .
    \end{split}
\end{equation}
We use the \texttt{dilute-lte}, \texttt{nebular} and \texttt{scatter} approximations for excitation, ionization and line treatment, respectively, as well as including the \texttt{recomb-nlte} He treatment \citep[as presented by][]{Boyle2017}, to capture NLTE excitation effects for \ion{He}{1}.
Our approximately LTE ionization and excitation treatments for all other ions are well-motivated as the phases we are probing are within the photospheric regime.

\begin{table}
    \renewcommand*{\arraystretch}{1.2}
    \centering
    \caption{
        \tardis\ model compositions.
    }
    \begin{tabular}{ll}
        \toprule
        Element     &Mass fraction      \\
        \midrule
        He          &0.50        \\
        O           &0.30        \\
        Si          &0.20        \\
        Ca          &$5 \times 10^{-6}$        \\
        \bottomrule
    \end{tabular}
    \label{tab:TARDIS abundance}
\end{table}

We present our model fits in Figure~\ref{fig:TARDIS fits}. 
The observations possess a number of prominent absorption features,
located at $\sim 5700$, 6100, 6400, 6800, 7400 and 8200\,\AA.
We find that we can reproduce the +4.2\,d spectrum with a relatively simple composition, made up of He, O, Si and Ca,
where the 6100\,\AA\ feature is produced by Si II, the 7400\,\AA\ feature by O I, the 8200\,\AA\ feature by Ca II, and all others (\ie, 5700, 6400 and 6800\,\AA) by He I. We over-produce the 5700\,\AA\ He absorption feature, and do not reproduce the
continuum blueward of $\lesssim 5400$\,\AA, but overall the fit to the data is good.

We note that the NLTE He treatment within \tardis\ is a simple, empirically derived approximation designed to account for the effects of recombination from \mbox{\ion{He}{2} $\xrightarrow{}$ \ion{He}{1}}. As such, it is possible that the estimated level populations within our \tardis\ simulation have deviated from the true level populations.
This possible issue has been noted before, and was proposed as the reason behind the disagreement between the relative strengths of He features in the case of
\SNxx{2019wxt} \citep{Agudo2023_SN2019wxt}. There, they manually altered the relative level populations of \ion{He}{1} to better match the observed relative strengths of the \ion{He}{1} features. Here we opt to not explore such variations, as our focus is on constraining the elements that dominate the composition in the line-forming region of the ejecta material of \thisSN.

Attempts to fit the observations in a more detailed manner than presented here should be approached with caution, given we have no reliable constraint on the true level of extinction. As a result, the true continuum of these observed spectra could be much bluer than what we present here (see Section~\ref{sec:host_extinction} for details on our extinction estimates), which would significantly alter the agreement of our models to the data.

Although our model composition is quite rudimentary, it aligns with our SN Ib classification, and not SN\,.Ia.
We are able to constrain the composition to be $\sim 80$\% He and O, and $\sim 20$\% Si (in the line-forming region). Our inner velocity estimates derived from this modeling (\mbox{$v_{\rm inner}^{+4.2\,{\rm d}} = 12500$\,km\,s$^{-1}$} and \mbox{$v_{\rm inner}^{+8.2\,{\rm d}} = 5000$\,km\,s$^{-1}$}) indicate the photosphere is receding quickly into the ejecta material, consistent with a small mass of ejected material.

\subsection{Volumetric Rates}
\label{sec:rates}

Here we present a preliminary volumetric rate estimate for \thisSN-like rapidly evolving stripped-envelope SNe (SESNe). We use the methodology described by \citet{Srivastav2022} for estimating rates of SNe Iax. We consider rapidly evolving SESNe detected by the ATLAS survey that occurred within a distance of 100 Mpc
during a 5-year window spanning 2017 September 21 and 2022 September 20 \citep{Srivastav2022}.
To estimate the recovery efficiency of \thisSN-like transients within 100 Mpc, we use the ATLAS survey simulator \citep{McBrien2021b}. We use Gaussian Processes interpolated ATLAS $c$ and $o$-band light curves of \thisSN, produced by interpolating the light curves using the public \texttt{extrabol} \citep{2023AAS...24110724T} code. These were then injected 10,000 times at a range of times, sky locations and redshift bins spanning up to D = 100 Mpc in the simulation. A simulated transient was considered to be recovered as a detection if it produced a minimum of 6 to 8 detections of $5\sigma$ (or greater) significance. Although difference detections in the ATLAS data stream are flagged as candidate transients if they produce 3 individual 
$5\sigma$ detections on any given night, this criterion is more realistic since human scanners will be confident about promoting candidates to the TNS if they have detections over  at least two distinct nights.

The volumetric rate is thus estimated using:
\begin{equation}
R = \frac{N}{\eta V T},
\end{equation}
where $T$ is the time duration of the mock survey, $N$ is the number of ultra-stripped SNe detected within the considered time duration, $\eta$ represents the recovery efficiency from the ATLAS survey simulator and $V$ is the volume probed within 100 Mpc. We consider $N = 3$, representing \SNxx{2019bkc} \citep{2020A&A...635A.186P,2020ApJ...889L...6C}, \SNxx{2019dge} \citep{2020ApJ...900...46Y} and \SNxx{2021agco} \citep{2023ApJ...959L..32Y}. 
The recovery efficiency obtained from the survey simulator is $\eta \approx 0.06 \pm 0.02$.

From the above, we estimate a rate of $R \approx 2.5^{+2.5}_{-1.4} \pm 0.9 \times 10^{-6}$ Mpc$^{-3}$ yr$^{-1}$ $h^3_{70}$ for rapidly evolving \thisSN-like SESNe, where $h_{70} = H_0 / 70$. The statistical uncertainty derives from $1\sigma$ Gaussian errors from single-sided upper and lower limits for Poisson statistics \citep{1986ApJ...303..336G} and the systematic uncertainty is based on the error on the recovery efficiency $\eta$. The above rate estimate for \thisSN-like SESNe accounts for $\sim 1-6\%$ of the CCSN and $\sim 5-20\%$ of the SESN rates computed by \citet{2021MNRAS.500.5142F}. We note here that $\sim 15\%$ of transients in the 100 Mpc ATLAS sample do not have a spectroscopic classification \citep{Srivastav2022}, and it is possible that the representation of \thisSN-like rapidly evolving SESNe is disproportionately higher within the unclassified sample. Nonetheless, at a few per cent of the CCSN rate, these transients clearly constitute a rare class of stellar explosions.

\subsection{Potential Evolutionary Route}

Using the parameters derived from our analysis we search for potential progenitor systems in the Binary Population and Spectral Synthesis data release \citep[BPASSv2.2.2;][]{bpass2017, bpass2018, hoki}. Assuming a solar metallicity (Z=0.02) we consider models with $M_{\rm H}<0.01$\,\msun\ and hydrogen mass fraction $X<0.001$ \citep{dessart12}. We use the inferred low ejecta mass from Section~\ref{sec:lightcurve_model} of  $ M_{\rm ej} <0.1$\,\msun\ and a low kinetic energy of $10^{50}$\,erg. Additionally, we apply a condition for explodability, commonly a mass threshold of the Oxygen Neon core ($>1.38$\,\msun), used to determine if a model is a candidate for core collapse. No models in BPASSv2.2.2 are found to match all of our conditions; however, slightly relaxing the condition to an ONe core mass $>1.30$\,\msun\ (one significant figure), we find 10 models that fit our criteria. When we include weights dependent on the Initial Mass Function \citep{kroupa01} as well as the binary fractions and period distributions \citep{moe17}, this corresponds to 28 systems per million solar masses. 

Grids of stellar evolution models such as BPASSv2.2.2 do not contain all possible observable outcomes, however we find candidate progenitor stars just at the threshold of explodability, motivating the observed rarity of \thisSN-like explosions. The progenitor of \thisSN\ is very likely a lower mass star and the initial masses of our 6 systems on the cusp of explodability, range from $7.5 - 9$\,\msun, which are some of the most common massive stars in the Universe. 
The low rate of \thisSN-like SNe and BPASS results are reconciled if potential progenitors of these SNe fail to explode most of the time, as their cores do not reach the necessary physical conditions. 
The 10 models with ONe core mass $> 1.3$\,\msun\ represent only 2.6 percent of the massive stars ($M_{\rm ZAMS}>7.5$\,\msun) that fit our stripping and ejecta mass criteria, and it is unlikely all of these would explode. This is in agreement with the calculated rates in Section \ref{sec:rates} of a few percent of the CCSN rate.

\section{Summary and Conclusions}

\label{sec:conclusion}

In this section we summarize the properties of \thisSN. 

\begin{enumerate}
    \item \thisSN\ shows a rapid rise ($<$ 4 rest-frame days), and initial decline from maximum light which settles to a radioactive tail 10 days after peak. Comparisons to ultra-stripped SNe and rapidly evolving supernovae shows that \thisSN\ is comparable to SN 2019bkc \citep{2020ApJ...889L...6C,2020A&A...635A.186P}. We consider radioactive nickel as the power source for \thisSN\,, finding that nickel alone cannot power both the peak and the tail of the light curve,  unlike SN 2019dge and SN 2019wxt \citep{ 2020ApJ...900...46Y,2023A&A...675A.201A}. An additional power source is required. 

   \item We consider several additional powering mechanisms and use agreement with the pseudo-bolometric light curve (Figure \ref{fig:phot}) and multi-band photometry (Figure \ref{fig:mosfit}) to select a preferred model. This analysis favors interaction with CSM material ($M_{\rm csm} \simeq 0.2$\,\msun, $r_{\rm csm} \simeq 63$\,AU), or powering from a central engine $(P_{\rm spin} \simeq 6$\,ms and $B_{\perp} \simeq 0.2 \times 10^{14}$\,G), to boost the initial luminosity before \thisSN\, settles to a \Co-tail. We note that signatures of interaction were not observed in the spectra, and suggest that any CSM envelope was swept up by the photosphere before our first spectroscopic observation with the spectroscopic narrow helium lines only becoming visible at late times.

    \item Through spectroscopic comparison we show \thisSN\ is similar to type Ib SNe and shows lines and line strengths similar to SN 2007C \citep{2014AJ....147...99M} and SN 2019wxt \citep{2023A&A...675A.201A}. Monte Carlo radiative transfer modeling with \tardis\ shows \thisSN\ has a composition dominated by He, O and Si. The spectroscopic evolution is not compatible with He shell detonation models. 

    \item A simulated ATLAS survey and estimate of the spectroscopic completeness of the ATLAS Volume Limited Survey (D $<$ 100 Mpc) yields a rate estimate of $R \approx 2.5^{+2.5}_{-1.4} \pm 0.9 \times 10^{-6}$ Mpc$^{-3}$ yr$^{-1}$ $h^3_{70}$. \thisSN-like transients could be as common as $\sim 1-6\%$ of the CCSN rate. Searching for potential progenitor stars in BPASS models, we propose that \thisSN-like events are the result of lower mass progenitors ($M_{\rm ZAMS} = 7.5 - 9$\,\msun) whose cores are at the threshold of explodability. The low observed rate of these SNe is then a result of the fact that only a few percent of these stars end with a core mass sufficient to result in core collapse.     
\end{enumerate}

We have shown \thisSN\ to be part of a small group of rapidly evolving SNe with a low ejecta mass ($M \simeq 0.07$\,\msun), and estimated a total nickel mass synthesized in the explosion ($M_{\rm Ni} \simeq 0.006$\,\msun). Furthermore, we find evidence in favor of an extra luminosity source in addition to the radioactive decay of \Ni. With our estimate of host galaxy extinction and significant Milky Way extinction in the line of sight we cannot reproduce the observe light curve with shock cooling emission, as favored by \citep{2024arXiv240308165D}, and instead we favor interaction with a detached CSM, or central engine energy injection, to boost the luminosity of \thisSN.

\section*{acknowledgments}
SJS, SS, KWS and DRY acknowledge funding from STFC Grants ST/Y001605/1, ST/X001253/1, ST/X006506/1 and ST/T000198/1. 
SJS acknowledges a Royal Society Research Professorship. 
MN is supported by the European Research Council (ERC) under the European Union's Horizon 2020 research and innovation programme (grant agreement No.~948381) and by UK Space Agency Grant No.~ST/Y000692/1. TWC acknowledges the Yushan Young Fellow Program by the Ministry of Education, Taiwan for the financial support. 
SY acknowledges the funding from the National Natural Science Foundation of China under Grant No. 12303046.
HFS is supported by the Eric and Wendy Schmidt A.I. in Science Fellowship.
Pan-STARRS is primarily funded to search for near-earth asteroids through NASA grants NNX08AR22G and NNX14AM74G. The Pan-STARRS science products for transient follow-up are made possible through the contributions of the University of Hawaii Institute for Astronomy and Queen’s University Belfast.
ATLAS is primarily funded through NASA grants NN12AR55G, 80NSSC18K0284, and 80NSSC18K1575. The ATLAS science products are provided by the University of Hawaii, Queen’s University Belfast, STScI, SAAO and Millennium Institute of Astrophysics in Chile. 
We thank Lulin staff H.-Y. Hsiao, W.-J. Hou, C.-S. Lin, H.-C. Lin, and J.-K. Guo for observations and data management.
Based on observations obtained at the international Gemini Observatory (under program ID GN-2023B-Q-125), a program of NSF NOIRLab, which is managed by the Association of Universities for Research in Astronomy (AURA) under a cooperative agreement with the U.S. National Science Foundation on behalf of the Gemini Observatory partnership: the U.S. National Science Foundation (United States), National Research Council (Canada), Agencia Nacional de Investigaci\'{o}n y Desarrollo (Chile), Ministerio de Ciencia, Tecnolog\'{i}a e Innovaci\'{o}n (Argentina), Minist\'{e}rio da Ci\^{e}ncia, Tecnologia, Inova\c{c}\~{o}es e Comunica\c{c}\~{o}es (Brazil), and Korea Astronomy and Space Science Institute (Republic of Korea). This work was enabled by observations made from the Gemini North telescope, located within the Maunakea Science Reserve and adjacent to the summit of Maunakea. We are grateful for the privilege of observing the Universe from a place that is unique in both its astronomical quality and its cultural significance.
Lasair is supported by the UKRI Science and Technology Facilities Council and is a collaboration between the University of Edinburgh (grant ST/N002512/1) and QUB (grant ST/N002520/1) within the LSST:UK Science Consortium. ZTF is supported by National Science Foundation grant AST-1440341 and a collaboration including Caltech, IPAC, the Weizmann Institute for Science, the Oskar Klein Center at Stockholm University, the University of Maryland, the University of Washington, Deutsches Elektronen-Synchrotron and Humboldt University, Los Alamos National Laboratories, the TANGO Consortium of Taiwan, the University of Wisconsin at Milwaukee, and Lawrence Berkeley National Laboratories.  

\facilities{Gemini:Gillett, Swift, PS1, PO:1.2m, Liverpool:2m}

\software{
        Astropy \citep{Astropy2013,Astropy2018,astropy2022},
        Numpy \citep{Harris2020}, 
        Matplotlib \citep{Hunter2007_matplotlib},
        Mosfit \citep{2018ApJS..236....6G},
        Hoki \citep{hoki},
        PSF \citep{2023ApJ...954L..28N},
        DRAGONS \citep{Labrie2023_DRAGONS, DRAGONS_zenodo}
}
\appendix
\section{MOSFiT Model Posterior}
\begin{figure*}[b]
    \centering 
    \includegraphics[width=0.9\linewidth]{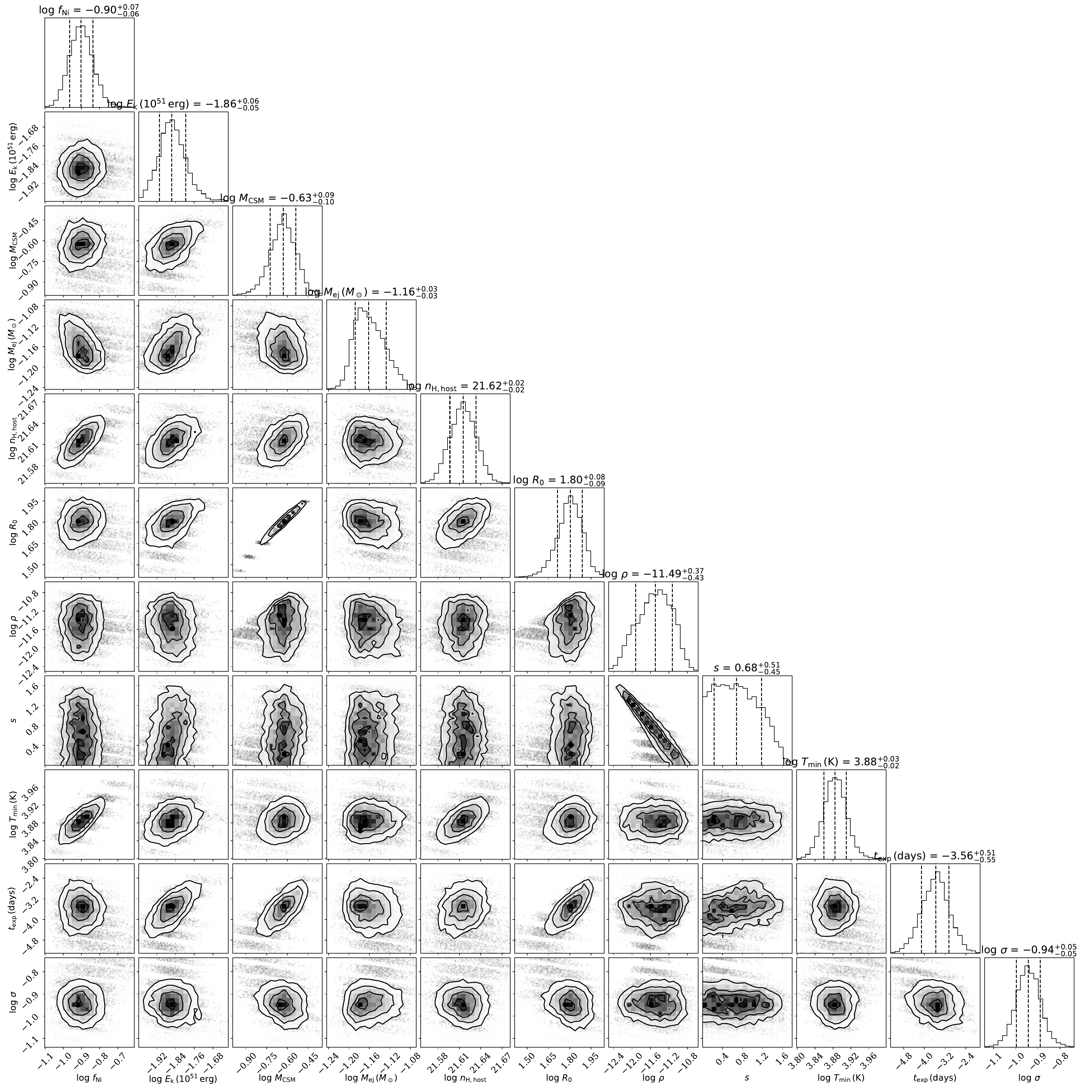}
    \caption{Physical parameter posterior distribution of the circumstellar material + nickel model. The important physical parameters are nickel fraction ($f_{\rm Ni}$), kinetic energy ($E_{\rm k}$), CSM mass ($M_{\rm csm}$), ejecta mass ($M_{\rm ej}$), the CSM radius ($R_{\rm 0}$) in units of AU, CSM density ($\rho$), and $t_{\rm exp}$ relative to the first photometric detection (MJD 60285.23) of \SNxx{2023zaw}.
    }
    \label{fig:csmni}
\end{figure*}

The MOSFiT model posterior for the \texttt{CSMNI} model is presented in Figure \ref{fig:csmni}. 

\bibliography{bib}
\pagebreak
\bibliographystyle{aasjournal}

\end{document}